\documentclass[11pt,a4paper]{article}
\pdfoutput=1

\usepackage[T1]{fontenc}
\usepackage[utf8]{inputenc}
\usepackage{lmodern}
\usepackage[margin=1in]{geometry}
\usepackage{amsmath,amsthm,amsfonts,amssymb}
\usepackage{enumerate}
\usepackage{algorithm}
\usepackage{algpseudocode}
\usepackage[hidelinks]{hyperref}
\hypersetup{pdftitle={Multiple-Choice Knapsack with Small Items},
  pdfauthor={Jakub Pawlewicz},
  pdfkeywords={multiple-choice knapsack, pseudopolynomial algorithms, proximity bounds, subset sums}}

\newtheorem{theorem}{Theorem}
\newtheorem{lemma}[theorem]{Lemma}
\newtheorem{corollary}[theorem]{Corollary}
\newtheorem{proposition}[theorem]{Proposition}
\theoremstyle{definition}
\newtheorem{definition}[theorem]{Definition}

\title{Multiple-Choice Knapsack with Small Items}
\author{Jakub Pawlewicz\\
  Institute of Informatics, University of Warsaw, Poland\\
  \texttt{pan@mimuw.edu.pl}\\
  \href{https://orcid.org/0000-0003-4670-9106}{ORCID: 0000-0003-4670-9106}}
\date{}

\newcommand{\solitems}[1]{S[#1]}

\newcommand{\vect}[1]{\boldsymbol{#1}}

\newcommand{\msum}[1]{\Sigma(#1)}
\newcommand{\mmax}[1]{\max(#1)}
\newcommand{\mallsums}[1]{\mathcal{S}(#1)}
\newcommand{\mallsumsplus}[1]{\mathcal{S}^*(#1)}

\newcommand{\mdiff}[3]{D^{#1}(#2,#3)}
\newcommand{\mdiffweights}[3]{D^{#1}_w(#2,#3)}

\newcommand{\Nset}{\mathbb{N}}
\newcommand{\Rset}{\mathbb{R}}

\newcommand{\Oh}{\mathcal{O}}
\newcommand{\tOh}{\tilde{\mathcal{O}}}

\begin{document}

\maketitle

\begin{abstract}
The Multiple-Choice Knapsack problem is a generalization of the Knapsack
problem in which items are partitioned into disjoint classes and exactly one
item must be selected from each class. Besides the standard dynamic program
running in time $\Oh(nc)$, where $n$ is the total number of items and $c$ is
the knapsack capacity, no other pseudopolynomial-time algorithm is currently
known. We
present an $\Oh(n+w_{\max}^4)$-time algorithm, where
$w_{\max}=\max_i (w_{in_i}-w_{i1})$ is the maximum weight range within a
class.
\end{abstract}

\section{Introduction}

The classical 0-1 Knapsack problem asks to choose a subset of items, each with
weight and profit, whose total weight does not exceed a capacity, while
maximizing the total profit. It is a cornerstone of combinatorial optimization
and a canonical NP-hard problem, motivating a broad toolkit ranging from
pseudopolynomial-time dynamic programs to approximation schemes and
fine-grained algorithmic results.

The Multiple-Choice Knapsack problem (MCKP) extends the 0-1 Knapsack problem
by partitioning items into disjoint classes and requiring exactly one item
from each class. It captures settings where each component must be chosen
from a menu of alternatives.

A closely related variant is the Selective MCKP, where at most one item per
class may be selected. It reduces to MCKP by adding a dummy item of weight
and price $0$ to each class, so we focus on the exact-choice formulation
without loss of generality.

MCKP appears in applications such as resource allocation, project selection,
and communication systems; see the survey by Szkaliczki~\cite{szkaliczki} for
a recent overview or the book by Kellerer et al.~\cite[Chapter 11]{kellerer-etal04}
for comprehensive coverage of Multiple-Choice Knapsack and related problems.
In a typical application, each class represents a decision that must be made
exactly once (e.g., selecting an implementation of a module, a mode of
operation for a user, or one project from a group of mutually exclusive
alternatives), and each item encodes a trade-off between resource consumption
(weight) and benefit (price). The knapsack capacity models the global resource
budget.
MCKP also arises in string algorithmics, e.g., in weighted pattern matching and
consensus problems on weighted sequences and profiles~\cite{kociumaka-etal18}.

\emph{Applications with small within-class ranges.}
Many communication and sensing problems lead to MCKP instances in which each
class has only a few discrete operating points, so the within-class weight
range $w_{in_i}-w_{i1}$ is naturally small. In
Section~\ref{sec:applications} we discuss representative examples including
bit allocation for crowdsourcing-based target tracking and discrete
transmit-power control in Zigbee sensor
networks~\cite{cao-brahma-varshney15,consolini-etal09}. We also note that the
same dynamic-programming machinery yields bounded-domain $(\max,+)$
convolution on consecutive ranges, which connects to deterministic network
calculus~\cite{cruz91,leboudec-thiran01}.

For the 0-1 Knapsack problem, classical pseudopolynomial-time dynamic programs
run in $\Oh(nc)$ for integral capacity $c$, or in $\Oh(nP)$ when
parameterized by the total price $P$, and yield approximation schemes. In the
small-weight regime, algorithms with running time
$\tOh(n+w_{\max}^2)$ have been presented for 0-1 Knapsack
~\cite{jin23,bringmann23}. For MCKP, the standard $\Oh(nc)$ dynamic program
remains essentially the only pseudopolynomial-time algorithm currently known;
see, e.g.,~\cite{dudzinski-walukiewicz87}.

Our main result is the following.
\begin{theorem}\label{thm:main}
    The Multiple-Choice Knapsack problem can be solved in time
    $\Oh(n+w_{\max}^4)$, where $n$ is the total number of items and $w_{\max}$
    is the maximum weight range within a class. Moreover, for any capacity
    $c$, the algorithm can output optimal objective values for the capacities
    $c,c+1,\ldots,c+w_{\max}$ within the same running time.
\end{theorem}
When item prices are integral and the maximum within-class price range
$p_{\max}$ is small, we also obtain a symmetric price-based variant running
in $\Oh(n+p_{\max}^4)$ (Section~\ref{sec:prices}).

At a high level, we first compute the upper hull of each class and
use it to construct a greedy anchor (Section~\ref{sec:greedy}). We then
prove a proximity lemma (Lemma~\ref{lem:proximity}) showing that there exists
an optimal solution for capacity $c$ that differs from the anchor in only
few classes, with bounded total weight increase and decrease. The proof
combines an exchange argument with an elementary subset-sum separation
lemma for two multisets whose positive subset sums are disjoint
(Section~\ref{sec:subset-sums}). Finally, we keep only the most promising
candidates for each possible weight change and run a dynamic program over the
total weight change within a window of size $\Oh(w_{\max}^2)$
(Section~\ref{sec:proximity}).

While MCKP is closely related to 0-1 Knapsack, standard reductions do not
preserve the structural regimes exploited by the fastest ``small item''
algorithms. In particular, normalizing by the lightest item in each class
creates upgrade items but also introduces a per-class restriction (at most one
upgrade per class), which breaks the unrestricted mixing of items that those
methods rely on. Our algorithm instead uses the greedy anchor as a reference point and applies
a proximity bound to search only within a bounded neighborhood whose size is
controlled by $w_{\max}$.

\section{Preliminaries}

We write $[k]=\{1,\ldots,k\}$ for $k\in\Nset$. A \emph{multiset} is a set with
multiplicities; $A'\subseteq A$ denotes a submultiset and $|A|$ is its size
counting multiplicities. For a multiset $A$ of integers we define
$\msum{A}$ as the sum of elements and $\mmax{A}$ as the maximum element. We
also use
\[
    \mallsums{A}=\{\msum{A'}\;:\;A'\subseteq A\},\qquad
    \mallsumsplus{A}=\mallsums{A}\setminus\{0\}.
\]

\section{Problem model}

The Multiple-Choice Knapsack problem is a generalization of the Knapsack
problem. Items are partitioned into classes, and exactly one item must be
selected from each class.

\subsection{Problem statement, notation, and assumptions}

Formally, the input consists of
\begin{itemize}
    \item the knapsack capacity: $c\in\Rset$,
    \item the number of classes: $r\in\Nset$,
    \item the number of items in each class: $n_i\in\Nset$ for $i\in[r]$,
    \item the weight and price of each item: $(w_{ij},p_{ij})\in\Rset$ for $i\in[r]$, $j\in[n_i]$.
\end{itemize}
A solution is an $r$-element sequence of indices $(j_1, \ldots, j_r)$,
denoted by $\vect{j}\in J$, where $J=[n_1]\times\ldots\times[n_r]$ is the
solution space. The weight and price of $\vect{j}$ are
\[
    w(\vect{j})=\sum_{i=1}^r w_{ij_i},\qquad
    p(\vect{j})=\sum_{i=1}^r p_{ij_i}.
\]
The goal is to maximize $p(\vect{j})$ subject to $w(\vect{j})\le c$ over
all $\vect{j}\in J$. If $\vect{j}$ maximizes $p(\vect{j})$ under the
constraint $w(\vect{j})\le c$, we say that $\vect{j}$ is an \emph{optimal
solution for capacity $c$}.

Each item is represented by a pair of indices $(i,j)$, $i\in[r]$,
$j\in[n_i]$. We denote the set of all items by
$S=\{(i,j):i\in[r],j\in[n_i]\}$ and the $i$-th class by
$S_i=\{(i,j):j\in[n_i]\}$. The items selected by a solution $\vect{j}$ are
$\solitems{\vect{j}}=\{(i,j_i):i\in[r]\}$.

An item $(w,p)$ is dominated within its class if there exists an item
$(w',p')$ in the same class with $w'\le w$ and $p'\ge p$. We assume that
such items are removed. Therefore, within each class items are arranged in
strictly increasing order of both weights and prices:
$w_{i1}<\cdots<w_{in_i}$ and $p_{i1}<\cdots<p_{in_i}$.
We assume that $\sum_{i=1}^r w_{i1}\le c$, since otherwise the instance is
infeasible, and that $c<\sum_{i=1}^r w_{in_i}$, since otherwise the
heaviest item in every class is optimal. Finally, item weights are assumed
to be integers. We denote
$w_{\max}=\max_{i=1}^r (w_{in_i}-w_{i1})$ and
$n=\sum_{i=1}^r n_i$.

\subsection{Relation to Selective Multiple-Choice Knapsack}

The selective version (Selective MCKP, SMCKP) allows picking at most one
item from each class, and classes may be skipped. It reduces to MCKP by
adding to every class a dummy item of weight and price $0$; picking this
item corresponds to skipping the class. Conversely, MCKP is a special case
of the selective version where choosing an item from every class is
required. Hence any algorithm for one variant applies to the other with
this simple reduction.

We can also reduce MCKP to SMCKP by normalizing each class with its
lightest item. Let $(w_{i1},p_{i1})$ be the lightest item in class $i$ and
set $c'=c-\sum_{i=1}^r w_{i1}$. For each $j>1$ define an upgrade item with
weight $w_{ij}-w_{i1}$ and price $p_{ij}-p_{i1}$. Selecting at most one
upgrade per class in the selective instance corresponds to choosing exactly
one item per class in the original instance, with the baseline items
providing the option of no upgrade.

\section{Greedy anchors}\label{sec:greedy}

We start by defining the greedy anchor used as the reference point for the
local search. The anchor is computed from the upper hull of every class, but the
final dynamic program in Section~\ref{sec:proximity} still considers the
original items. This distinction is important: points below the upper hull
need not be removed from the instance, but they are dominated by the
piecewise-linear upper envelope used in the proof of optimality of the
anchor.

\begin{definition}
    For a class represented by upper-hull vertices, and for every $j\ge 2$,
    define the weight increment, price increment, and slope of the upgrade
    from item $j-1$ to item $j$ by
    \[
        \Delta^w_{i,j}=w_{ij}-w_{i(j-1)},\qquad
        \Delta^p_{i,j}=p_{ij}-p_{i(j-1)},\qquad
        \rho_{i,j}=\frac{\Delta^p_{i,j}}{\Delta^w_{i,j}}.
    \]
    An instance is \emph{concave} if
    $\rho_{i,j}\ge \rho_{i,j+1}$ for all $i\in[r]$ and
    $j\in\{2,\ldots,n_i-1\}$, i.e., marginal slopes within each class are
    non-increasing.
\end{definition}

For each class of an arbitrary instance, we compute the vertices of the
upper convex hull of the points $(w_{ij},p_{ij})$. The resulting
piecewise-linear envelope is concave as a function of weight; this is why we
use the term ``concave'' for the corresponding upper-hull subinstance. Since
items in a class are already sorted by weight, the upper hull of class $i$
can be computed in $\Oh(n_i)$ time by the standard monotone stack procedure,
and hence all upper hulls can be computed in $\Oh(n)$ time.

The greedy procedure starts with the lightest upper-hull item in each class.
It then processes upper-hull upgrades in non-increasing order of slope. Once
an upgrade of the current maximum remaining slope does not fit into the
remaining budget, the procedure may still take other upgrades of the same
slope that fit, but it stops before taking any upgrade of smaller slope. This
is the intended prefix-by-slope rule; in particular, the greedy anchor is not
obtained by skipping an overweight high-slope upgrade and then continuing
with lower-slope upgrades.

Let $w_0=\sum_{i=1}^r w_{i1}$ and $B=c-w_0$ be the upgrade budget. A
linear-time weighted selection on the slopes~\cite{selection} finds a
threshold $\tau$ such that all upgrades with slope larger than $\tau$ fit,
while not all upgrades with slope at least $\tau$ fit. We then scan the
classes once to take all larger-slope upgrades, and once more to take
threshold-slope upgrades while they fit. The resulting solution is the
\emph{greedy anchor} for capacity $c$.

\begin{algorithm}[htbp]
\caption{Greedy anchor for an upper-hull instance.}
\label{alg:greedy}
\begin{algorithmic}[1]
\Require Upper-hull classes $S_1,\ldots,S_r$, capacity $c$
\State $j_i \gets 1$ for all $i$, $w_0\gets \sum_{i=1}^r w_{i1}$
\If{$w_0 \ge c$}
    \State \Return $\vect{g}=(j_1,\ldots,j_r)$
\EndIf
\State $B\gets c-w_0$
\State Find a threshold $\tau$ such that
    $\sum_{\rho_{i,j}>\tau} \Delta^w_{i,j}\le B < \sum_{\rho_{i,j}\ge\tau} \Delta^w_{i,j}$
\For{each class $i$}
    \While{$j_i+1\le n_i$ and $\rho_{i,j_i+1}>\tau$}
        \State $j_i\gets j_i+1$, $B\gets B-\Delta^w_{i,j_i}$
    \EndWhile
\EndFor
\For{each class $i$}
    \While{$j_i+1\le n_i$ and $\rho_{i,j_i+1}=\tau$ and $\Delta^w_{i,j_i+1}\le B$}
        \State $j_i\gets j_i+1$, $B\gets B-\Delta^w_{i,j_i}$
    \EndWhile
\EndFor
\State \Return $\vect{g}=(j_1,\ldots,j_r)$
\end{algorithmic}
\end{algorithm}

\begin{proposition}[Greedy anchor]\label{prop:greedy}
        The greedy anchor can be computed in $\Oh(n)$ time. If Algorithm~\ref{alg:greedy}
    returns $\vect{g}$ and $c_g=w(\vect{g})$, then $\vect{g}$ is an optimal
    solution for capacity $c_g$ in the original instance. Moreover,
    $c-c_g<w_{\max}$.
\end{proposition}
\begin{proof}
    The running-time bound follows from the linear-time upper-hull
    computation, the linear-time weighted selection of the threshold slope,
    and the two scans over the upper-hull upgrades.

    For optimality, consider for every class $i$ the piecewise-linear upper
    envelope $f_i$ obtained by interpolating between consecutive upper-hull
    vertices. Every original item $(i,j)$ satisfies
    $p_{ij}\le f_i(w_{ij})$, with equality for upper-hull vertices. Hence any
    feasible MCKP solution of weight at most $c_g$ is feasible for the
    relaxation
    \newpage
    \[
        \max \sum_{i=1}^r f_i(x_i)
        \quad\text{subject to}\quad
        \sum_{i=1}^r x_i\le c_g,
        \quad w_{i1}\le x_i\le w_{in_i},
    \]
    and its price is at most the relaxation value.

    The functions $f_i$ are concave, and their linear pieces have slopes
    $\rho_{i,j}$. A standard exchange argument proves that a prefix of pieces
    in non-increasing slope order is optimal for the relaxation: if an
    allocation uses a positive amount of a lower-slope piece while a
    higher-slope piece is not full, shifting a sufficiently small amount of
    weight from the former to the latter preserves feasibility and does not
    decrease the objective (and increases it for a strict slope inequality).
    Algorithm~\ref{alg:greedy} produces such a prefix at total weight $c_g$,
    with arbitrary tie-breaking among pieces of slope $\tau$. Therefore the
    relaxation optimum at capacity $c_g$ equals $p(\vect{g})$. Since
    $\vect{g}$ itself is an original feasible solution, it is optimal for the
    original instance at capacity $c_g$.

    Finally, after the second scan no remaining upgrade of slope $\tau$ fits
    into the residual budget, and all lower-slope upgrades are deliberately
    ignored. The threshold exists because the all-heaviest solution is assumed
    to exceed $c$. Thus the residual budget $c-c_g$ is smaller than the weight
    of some remaining upper-hull upgrade, and every such upgrade has weight at
    most $w_{\max}$.
\end{proof}

\section{A subset-sum separation lemma}\label{sec:subset-sums}

To control how far an optimal solution can deviate from the greedy anchor,
we need a bound on cancellations: when some classes switch to lighter items
and others to heavier items, the corresponding weight differences form two
multisets. If these multisets shared a nonzero common subset sum, we could
exchange those classes and contradict the minimality argument in the
proximity proof. The following elementary pigeonhole lemma is the same kind
of subset-sum separation used by Polak, Rohwedder, and Węgrzycki for
0-1 Knapsack~\cite[Lemma 2.1]{polak-etal21}; we include a self-contained
variant for the notation used here. Throughout this section, all multisets
consist of positive integers.

\begin{theorem}\label{thm:AB-sums}
    Assume we have two multisets $A$ and $B$ of positive integers with
    $\msum{A}=\msum{B}=s$ and
    $\mallsumsplus{A}\cap\mallsumsplus{B}=\{s\}$.
    Then $|A|\le \mmax{B}$ and $|B|\le \mmax{A}$.
    Thus for $w$ such that $\mmax{A},\mmax{B}\le w$, we have $s\le w(w-1)$.
\end{theorem}
\begin{proof}
    Let $b_{\max}=\mmax{B}$, $n=|A|$, $m=|B|$.
    Enumerate all elements of $A$ and $B$ in no particular order:
    $A=\{a_1, \ldots, a_n\}$, $B=\{b_1, \ldots, b_m\}$.
    Define prefix sums $P_A(i)=\sum_{k\le i} a_k$ and $P_B(j)=\sum_{k\le j} b_k$
    for $0\le i\le n$ and $0\le j\le m$.
    For each $i$ with $1\le i < n$, define
    $j^*(i)=\max\{j:P_B(j)\le P_A(i)\}$.

    Observe that for $1\le i<n$ we have $j^*(i)<m$, hence
    $b_{j^*(i)+1}$ is defined and
    $0\le \Delta(i)=P_A(i)-P_B(j^*(i))<b_{j^*(i)+1}\le b_{\max}$.
    If $\Delta(i)=0$ then $P_A(i)=P_B(j^*(i))$ gives a nontrivial common sum,
    contradicting $\mallsumsplus{A}\cap\mallsumsplus{B}=\{s\}$.
    Therefore $\Delta(i)\in[1,b_{\max}-1]$.

    If $\Delta(i_1)=\Delta(i_2)$ for some $1\le i_1<i_2<n$, then
    $P_A(i_2)-P_A(i_1)=P_B(j^*(i_2))-P_B(j^*(i_1))$, which again contradicts
    the assumption $\mallsumsplus{A}\cap\mallsumsplus{B}=\{s\}$.
    Therefore the values $\Delta(i)$ are distinct for all $1\le i < n$, which
    implies $n-1\le b_{\max}-1$ and proves $|A|\le \mmax{B}$. By symmetry we
    obtain $|B|\le \mmax{A}$.

    Now assume $\mmax{A}, \mmax{B}\le w$. If $w\in B$ then $\mmax{A}<w$ and
    $A$ contains at most $w$ elements of value at most $w-1$. If $w\notin B$
    then $A$ contains at most $w-1$ elements of value at most $w$. Either way
    $s=\msum{A}\le w(w-1)$.
\end{proof}

\begin{corollary}\label{cor:AB-sums}
    Assume we have two multisets $A$ and $B$ of positive integers with
    $\msum{B}-\msum{A}=d>0$ such that for some $w$ we have
    $\mmax{A},\mmax{B},d\le w$.
    If $\mallsumsplus{A}\cap\mallsumsplus{B}=\emptyset$, then
    $|A|+1,|B|\le w$ and
    $\msum{B}=\msum{A}+d\le w(w-1)$.
\end{corollary}
\begin{proof}
    Apply Theorem~\ref{thm:AB-sums} to $A\cup\{d\}$ and $B$. Put
    $s=\msum{A}+d=\msum{B}$. We only need to verify that the positive subset
    sums of $A\cup\{d\}$ and $B$ intersect only at $s$.

    Suppose $\msum{A'}=\msum{B'}$ for some $A'\subseteq A\cup\{d\}$ and
    $B'\subseteq B$. If $d\notin A'$, then the common sum lies in
    $\mallsumsplus{A}\cap\mallsumsplus{B}$ unless it is zero; hence
    $A'=B'=\emptyset$.

    Now let $d\in A'$. Consider the complements
    $\bar{A}=A\cup\{d\}\setminus A'$ and $\bar{B}=B\setminus B'$.
    Since $\msum{A}+d=\msum{B}$ and $\msum{A'}=\msum{B'}$, we have
    $\msum{\bar{A}}=\msum{\bar{B}}$. Moreover $\bar{A}\subseteq A$ and
    $\bar{B}\subseteq B$. If this common sum were positive, it would belong to
    $\mallsumsplus{A}\cap\mallsumsplus{B}$, impossible. Thus
    $\bar{A}=\bar{B}=\emptyset$, and so $A'=A\cup\{d\}$ and $B'=B$.
\end{proof}

\section{Proximity}\label{sec:proximity}

Intuitively, the greedy anchor $\vect{g}$ is already close to optimal for
capacity $c$: it may leave some unused slack, but the slack is bounded by
$w_{\max}$. Any optimal solution must therefore differ from $\vect{g}$ in only
a small number of classes, and large cancellations between weight decreases
and increases would allow an exchange that improves $\vect{g}$. We turn this
observation into an algorithm in three steps. First, we show that some optimal
solution lies within a bounded neighborhood of the anchor. Next, we retain only
$\Oh(w_{\max}^2)$ candidate changes. Finally, we solve the resulting reduced
instance by dynamic programming.

\subsection{Optimal solutions near an anchor}

We begin by introducing notation for the differences between two solutions.

\begin{definition}
    For two solutions $\vect{g}, \vect{j}\in J$ we define the following sets
    of class indices representing the differences between these solutions:
    $\mdiff{\bullet}{\vect{g}}{\vect{j}}=\{i\in[r]\;:\;w_{ig_i}\bullet w_{ij_i}\}$
    where $\bullet$ is a comparison relation: $\bullet\in\{\neq,>,<\}$.
    Thus $\mdiff{\neq}{\vect{g}}{\vect{j}}$ denotes the set of class indices
    where solutions differ, $\mdiff{<}{\vect{g}}{\vect{j}}$ those where the
    second solution has a heavier item, and $\mdiff{>}{\vect{g}}{\vect{j}}$
    those where the second solution has a lighter item.

    Additionally, we define the multiset of weight differences
    $\mdiffweights{\bullet}{\vect{g}}{\vect{j}}=\{|w_{ij_i}-w_{ig_i}|\;:\;i\in\mdiff{\bullet}{\vect{g}}{\vect{j}}\}$
    for $\bullet\in\{>,<\}$.
    The multiset $\mdiffweights{>}{\vect{g}}{\vect{j}}$ (respectively,
    $\mdiffweights{<}{\vect{g}}{\vect{j}}$) contains the weight differences by
    which the item weight was decreased (increased) between solutions
    $\vect{g}$ and $\vect{j}$.
\end{definition}

\begin{lemma}\label{lem:proximity}
    Let $\vect{g}$ be an optimal solution for capacity $c'$ with
    $w(\vect{g})=c'$ and $0\le c-c'\le w_{\max}$.
    Then there exists an optimal solution $\vect{j}$ for capacity $c$ such that
    \begin{itemize}
        \item $|\mdiff{>}{\vect{g}}{\vect{j}}|< w_{\max}$ and $\msum{\mdiffweights{>}{\vect{g}}{\vect{j}}}<w_{\max}(w_{\max}-1)$,
        \item $|\mdiff{<}{\vect{g}}{\vect{j}}|\le w_{\max}$ and $\msum{\mdiffweights{<}{\vect{g}}{\vect{j}}}\le w_{\max}(w_{\max}-1)$.
    \end{itemize}
\end{lemma}
\begin{proof}
    Let $\vect{j}$ be an optimal solution for capacity $c$ that minimizes the
    size of $\mdiff{\neq}{\vect{g}}{\vect{j}}$.
    If $w(\vect{j})\le c'$, then $\vect{j}$ is feasible for capacity $c'$.
    Since $\vect{g}$ is optimal for capacity $c'$, we have
    $p(\vect{g})\ge p(\vect{j})$. But $\vect{g}$ is feasible for capacity $c$
    and $\vect{j}$ is optimal for capacity $c$, so $\vect{g}$ is also optimal
    for capacity $c$. Replacing $\vect{j}$ by $\vect{g}$ gives empty difference
    sets, which satisfy the stated bounds. Thus assume from now on that
    $w(\vect{j})>c'$.

    Let $d=w(\vect{j})-w(\vect{g})$. Then
    \[
        \msum{\mdiffweights{>}{\vect{g}}{\vect{j}}}+d
        =\msum{\mdiffweights{<}{\vect{g}}{\vect{j}}},
    \]
    and $0<d\le c-c'\le w_{\max}$.

    We claim that
    \[
        \mallsumsplus{\mdiffweights{>}{\vect{g}}{\vect{j}}}\cap
        \mallsumsplus{\mdiffweights{<}{\vect{g}}{\vect{j}}}=\emptyset .
    \]
    Suppose otherwise. Then there exist nonempty sets
    $I^>\subseteq\mdiff{>}{\vect{g}}{\vect{j}}$ and
    $I^<\subseteq\mdiff{<}{\vect{g}}{\vect{j}}$ such that
    \[
        \sum_{i\in I^>}(w_{ig_i}-w_{ij_i})
        =\sum_{i\in I^<}(w_{ij_i}-w_{ig_i}).
    \]
    Let $I=I^>\cup I^<$. Then $\sum_{i\in I}w_{ig_i}=\sum_{i\in I}w_{ij_i}$.
    Since $\vect{g}$ is optimal for capacity $c'$ in the original instance, we
    must have $\sum_{i\in I}p_{ig_i}\ge\sum_{i\in I}p_{ij_i}$; otherwise
    swapping the classes of $I$ in $\vect{g}$ to the items of $\vect{j}$ would
    preserve weight and strictly increase price.

    Now form $\vect{j}^*$ from $\vect{j}$ by replacing, for every $i\in I$, the
    item of $\vect{j}$ by the item of $\vect{g}$. The total weight is preserved
    and the price does not decrease:
    $w(\vect{j}^*)=w(\vect{j})$ and $p(\vect{j}^*)\ge p(\vect{j})$. Hence
    $\vect{j}^*$ is also optimal for capacity $c$, but it differs from
    $\vect{g}$ in fewer classes than $\vect{j}$ does, contradicting the choice
    of $\vect{j}$.

    Therefore the multisets
    $A=\mdiffweights{>}{\vect{g}}{\vect{j}}$ and
    $B=\mdiffweights{<}{\vect{g}}{\vect{j}}$ satisfy the assumptions of
    Corollary~\ref{cor:AB-sums} with $w=w_{\max}$. The stated bounds follow.
\end{proof}

\subsection{Candidate pruning}

Given an anchor $\vect{g}$, exchanging the selected item in class $i$ with
item $(i,j)$ changes the weight and price by
\[
    \omega_{ij}=w_{ij}-w_{ig_i},\qquad
    \varphi_{ij}=p_{ij}-p_{ig_i}.
\]
The notation $\omega_{ij},\varphi_{ij}$ is used only for changes relative to
$\vect{g}$; it is distinct from the upper-hull increments
$\Delta^w_{i,j},\Delta^p_{i,j}$ used in Section~\ref{sec:greedy}. For every
nonzero integer
\[
    \delta\in\{-w_{\max},\ldots,-1,1,\ldots,w_{\max}\},
    \qquad
    G_\delta=\{(i,j):\omega_{ij}=\delta\}.
\]
We retain, in each group $G_\delta$, the $2w_{\max}$ items with largest price
change $\varphi_{ij}$ (or all items if the group is smaller). Let
$S_{\vect{g}}$ be the union of retained items. Then
$|S_{\vect{g}}|\le 4w_{\max}^2$.

The factor $2w_{\max}$ is chosen to dominate the total proximity bound
$2w_{\max}-1$, not the number of changes in one weight-change group. This
extra slack is needed because, when repairing a fixed group $G_\delta$, some
of its most profitable retained candidates may lie in classes already changed
through other groups. Such candidates are unusable under the ``one item per
class'' constraint.

\begin{lemma}[Candidate pruning]\label{lem:candidate-pruning}
    Among the optimal solutions for capacity $c$ guaranteed by
    Lemma~\ref{lem:proximity}, there is one whose every changed item belongs
    to $S_{\vect{g}}$.
\end{lemma}
\begin{proof}
    Choose, among the optimal solutions satisfying Lemma~\ref{lem:proximity},
    a solution $\vect{j}$ that maximizes the number of changed classes whose
    corresponding item is retained. Put
    $T=|\mdiff{\neq}{\vect{g}}{\vect{j}}|$. By Lemma~\ref{lem:proximity},
    $T\le 2w_{\max}-1$.

    Suppose that, for some group $G_\delta$, the solution $\vect{j}$ uses an
    unretained item. Let $q$ be the number of changed classes of $\vect{j}$ in
    $G_\delta$. Since weights are strictly ordered inside every class, each
    class contributes at most one candidate to a fixed group $G_\delta$. Hence
    each of the $T-q$ classes changed by $\vect{j}$ in a different group can
    forbid at most one retained candidate of $G_\delta$. Among the
    $2w_{\max}$ retained candidates of $G_\delta$, at least
    $2w_{\max}-(T-q)>q$ are therefore not forbidden. Choose $q$ such retained
    candidates with largest price changes. Every retained candidate has price
    change at least that of every unretained candidate in the same group, so
    the total price change of the chosen retained candidates is at least the
    total price change of the $q$ items used by $\vect{j}$ in $G_\delta$.

    Replacing the $G_\delta$ changes of $\vect{j}$ by these chosen retained
    changes preserves class feasibility, preserves the total weight change
    because all involved items have the same $\omega_{ij}=\delta$, and does
    not decrease the price. The proximity bounds are preserved as well. The
    resulting solution is optimal and has more retained changed items than
    $\vect{j}$, a contradiction.
\end{proof}

\subsection{Dynamic programming around an anchor}

Let $W=w_{\max}(w_{\max}-1)$. For each class $i$ define
\[
    \mathcal{C}_i=\{(0,0)\}\cup
    \{(\omega_{ij},\varphi_{ij}):(i,j)\in S_{\vect{g}}\}.
\]
Only classes with $\mathcal{C}_i\ne\{(0,0)\}$ are processed. For these
classes we compute a rolling dynamic program over total weight change
$x\in[-W,W]$. After processing some classes, $\mathrm{DP}(x)$ stores the
maximum total price change achievable with total weight change $x$; unreachable
states have value $-\infty$. The transition for a processed class $i$ is
\[
    \mathrm{DP}'(x+\omega)=
    \max\{\mathrm{DP}'(x+\omega),\mathrm{DP}(x)+\varphi\}
    \qquad\text{for }(\omega,\varphi)\in\mathcal{C}_i.
\]

\begin{algorithm}[htbp]
\caption{Dynamic program around an anchor solution.}
\label{alg:dp}
\begin{algorithmic}[1]
\Require Classes $S_1,\ldots,S_r$, anchor solution $\vect{g}$, capacity $c$
\State $W\gets w_{\max}(w_{\max}-1)$, $B\gets c-w(\vect{g})$
\State For each nonzero $\delta\in[-w_{\max},w_{\max}]$, keep the
    $2w_{\max}$ items of $G_\delta$ with largest $\varphi_{ij}$
\State For each class $i$ build
    $\mathcal{C}_i=\{(0,0)\}\cup\{(\omega_{ij},\varphi_{ij}):(i,j)\in S_{\vect{g}}\}$
\State Initialize $\mathrm{DP}(x)=-\infty$ for $x\in[-W,W]$ and set $\mathrm{DP}(0)=0$
\For{each class $i$ with $\mathcal{C}_i\ne\{(0,0)\}$}
    \State Initialize $\mathrm{DP}'(x)=-\infty$ for all $x\in[-W,W]$
    \For{each $x\in[-W,W]$}
        \For{each $(\omega,\varphi)\in\mathcal{C}_i$}
            \If{$x+\omega\in[-W,W]$}
                \State $\mathrm{DP}'(x+\omega)\gets \max\{\mathrm{DP}'(x+\omega),\,\mathrm{DP}(x)+\varphi\}$
            \EndIf
        \EndFor
    \EndFor
    \State $\mathrm{DP}\gets \mathrm{DP}'$
\EndFor
\State \Return $p(\vect{g})+\max\{\mathrm{DP}(x):x\in[-W,W],\ x\le B\}$
\end{algorithmic}
\end{algorithm}

\begin{proposition}[DP correctness and running time]\label{prop:dp}
    Let $\vect{g}$ be an optimal solution for capacity $c'=w(\vect{g})$ with
    $0\le c-c'\le w_{\max}$. Algorithm~\ref{alg:dp} outputs the optimal value
    for capacity $c$ in the original instance. Its running time is
    $\Oh(n+w_{\max}^4)$ and its working space is $\Oh(w_{\max}^2)$.
\end{proposition}
\begin{proof}
    By Lemma~\ref{lem:proximity} and Lemma~\ref{lem:candidate-pruning}, there
    exists an optimal solution for capacity $c$ whose changed items are all in
    $S_{\vect{g}}$ and whose total weight change lies in $[-W,W]$. The dynamic
    program enumerates exactly all choices of at most one retained change per
    class, grouped by their total weight change. Therefore the final maximum
    over $x\le c-w(\vect{g})$ contains the value of an optimal solution and
    cannot exceed the optimum, since every DP state corresponds to a feasible
    set of class changes.

    The pruning step is implemented by linear-time selection in each weight
    change group, and the groups together contain $O(n)$ items. The number of
    retained candidates is $M=|S_{\vect{g}}|\le 4w_{\max}^2$, so the number of
    active classes is at most $M$. The DP uses two arrays indexed by
    $x\in[-W,W]$, hence $\Oh(W)=\Oh(w_{\max}^2)$ space. Processing a class $i$
    costs $\Oh(W|\mathcal{C}_i|)$ time, and over all active classes
    $\sum_i |\mathcal{C}_i|\le 2M$. Thus the DP time is
    $\Oh(WM)=\Oh(w_{\max}^4)$, and the total time is
    $\Oh(n+w_{\max}^4)$.
\end{proof}

\begin{corollary}[Consecutive capacities]\label{cor:window}
    Fix an anchor $\vect{h}$ that is optimal for capacity
    $c_h=w(\vect{h})$. One run of the dynamic program around $\vect{h}$ gives
    optimal values for every capacity
    $\hat c\in\{c_h,c_h+1,\ldots,c_h+w_{\max}\}$.
\end{corollary}
\begin{proof}
    The DP table computed around $\vect{h}$ is independent of the final
    residual capacity. By Proposition~\ref{prop:dp}, for every such
    $\hat c$ the optimal value is
    \[
        p(\vect{h})+
        \max\{\mathrm{DP}_{\vect{h}}(x):x\in[-W,W],\ x\le \hat c-c_h\}.
    \]
    After computing prefix maxima of $\mathrm{DP}_{\vect{h}}(x)$ over the
    ordered range $x=-W,-W+1,\ldots,W$, all $w_{\max}+1$ values are read off
    in $\Oh(w_{\max})$ additional time.
\end{proof}

We can now prove Theorem~\ref{thm:main}. Compute the upper hulls and the
greedy anchor $\vect{g}$ for the requested capacity $c$. By
Proposition~\ref{prop:greedy}, $\vect{g}$ is optimal for
$c_g=w(\vect{g})$ in the original instance and $c-c_g<w_{\max}$. Therefore
Proposition~\ref{prop:dp} gives the optimal value for capacity $c$.

For the consecutive-capacity statement, Corollary~\ref{cor:window} applied to
$\vect{g}$ gives optimal values for
$c_g,c_g+1,\ldots,c_g+w_{\max}$. If these already cover
$c,c+1,\ldots,c+w_{\max}$, we are done. Otherwise, continue the greedy sweep
by one more upper-hull upgrade and let $\vect{g}'$ be the resulting anchor.
The next upgrade has weight at most $w_{\max}$, hence
\[
    c < w(\vect{g}') \le w(\vect{g})+w_{\max}.
\]
Corollary~\ref{cor:window} applied to $\vect{g}'$ gives optimal values for
$w(\vect{g}'),w(\vect{g}')+1,\ldots,w(\vect{g}')+w_{\max}$. The two intervals
cover all capacities $c,c+1,\ldots,c+w_{\max}$. Running the dynamic program
once or twice changes the running time only by a constant factor, so the
total time remains $\Oh(n+w_{\max}^4)$.

\section{Small item prices}\label{sec:prices}

Assume prices are integers and define
$p_{\max}=\max_{i\in[r]}(p_{in_i}-p_{i1})$. The price-based variant mirrors
the weight-based approach, but with roles reversed. First, the greedy
anchor is constructed by following the steepest price-per-weight slopes,
and it leaves a bounded \emph{price slack} rather than a weight slack. That
is, once the greedy process stops, any remaining upgrade would exceed the
capacity, so the additional price that could be gained by further upgrades is
limited by the largest price step in any class. Second,
we group candidate upgrades by their price change $\Delta p$ (instead of
weight change) and keep, for each $\Delta p$, the $2p_{\max}$ items with the
smallest weight increase. This guarantees that the multiset of price changes
used by an optimal solution has total price change at most
$P=p_{\max}(p_{\max}-1)$.

We then run a dynamic program indexed by total price change
$y\in[-P,P]$, where the state stores the minimum weight increase achievable
with those upgrades. This is the dual of the weight-based DP, which stores
maximum price increase for a given weight change. The same counting argument
gives $\Oh(p_{\max}^2)$ candidates and total running time
$\Oh(n+p_{\max}^4)$.

\section{Applications}\label{sec:applications}

The running time of the standard dynamic program for MCKP is $\Oh(nc)$ for
integral capacity $c$, which can be prohibitive when the global budget is
large. Our bound $\Oh(n+w_{\max}^4)$ replaces this dependence on the global
budget by $w_{\max}=\max_i (w_{in_i}-w_{i1})$, the largest \emph{within-class}
weight range. This regime is natural in engineered systems where each decision
offers only a few discrete operating points. The same bounded-window dynamic
program can also be viewed as a routine for bounded-domain $(\max,+)$
convolution on consecutive output ranges.

\subsection{\texorpdfstring{Bounded-domain $(\max,+)$ convolution}{Bounded-domain (max,+) convolution}}

Our algorithm can be interpreted as evaluating a bounded-domain
$(\max,+)$ convolution.
Given functions
$f_1,\ldots,f_k : \{0,\ldots,d\}\rightarrow\mathbb{R}$,
define

\[
F(x)=\max_{x_1+\cdots+x_k=x}\sum_{i=1}^k f_i(x_i).
\]

A straightforward dynamic programming approach computes
$F(0),\ldots,F(d)$ in $\Oh(kd^2)$ time.
Our algorithm implies that values of $F(x)$ on $d+1$ consecutive
points can be computed in time $\Oh(dk+d^4)$.
Consequently, values on an interval of length $s$ can be obtained in
$\Oh(s(k+d^3))$ time. This improves over the classical $\Oh(sk d)$ dynamic
programming approach when $k \gg d^2$.

\subsection{Deterministic network calculus}

Deterministic network calculus models network elements by service
curves and combines them using min-plus convolution.
For servers with service curves
$\beta_1,\ldots,\beta_k$, the end-to-end service curve is

\[
(\beta_1\otimes\cdots\otimes\beta_k)(t)
=\inf_{t_1+\cdots+t_k=t}\sum_{i=1}^k \beta_i(t_i).
\]

Such convolutions are a central computational primitive in DNC analyses
of delay and backlog bounds~\cite{cruz91,leboudec-thiran01}.

After a sign change this operation becomes a $(\max,+)$ convolution.
When service curves are discretized on a bounded time horizon
$t\in\{0,\ldots,d\}$, evaluating end-to-end guarantees reduces to
computing such bounded-domain convolutions.

In many scenarios one is interested only in values of the convolution
within a limited time window (for example when testing deadline
feasibility or during parametric search for delay bounds).
In such cases the windowed computation enabled by our algorithm can
reduce the running time when the number of network elements is large
but the discretization horizon is moderate.

\subsection{Bit allocation under a global communication budget}

Cao, Brahma, and Varshney study incentive-compatible target tracking via
crowdsourcing under a global communication budget. In their formulation, the
fusion center chooses for each sensor a number of bits to purchase subject to
a total bit budget, which they cast as an MCKP and solve by dynamic
programming in $\Oh(NM)$ time for $N$ sensors and budget $M$~\cite{cao-brahma-varshney15}.
When each sensor has only a small set of admissible bit depths (e.g., due to
quantizer resolution or protocol constraints), the per-sensor bit range is
small, so $w_{\max}$ is bounded independently of $M$ and the MCKP subproblem
can be solved in time $\Oh(n+w_{\max}^4)$. This is particularly beneficial
when the allocation must be recomputed repeatedly, e.g., across time steps in
tracking or as part of a larger mechanism.

\subsection{Transmit-power allocation with discrete power levels}

Consolini, Medagliani, and Ferrari consider transmit-power allocation in Zigbee
wireless sensor networks with a finite total power budget. They define a
discrete optimization problem in which each node selects one transmit power
from a small discrete set under a global power constraint, observe that it
corresponds to the multiple-choice knapsack problem, and solve it using an
off-the-shelf solver (MOSEK), noting that computation time grows quickly with
the number of sensors~\cite{consolini-etal09}. Here the set of available power
levels per node is typically small, so after discretization the within-class
range $w_{\max}$ is small even if the total budget is large. In this setting,
an $\Oh(n+w_{\max}^4)$-time algorithm gives an exact alternative whose running
time is controlled by the discretization granularity rather than by the global
budget.

\section{Conclusions}

We presented a pseudopolynomial-time algorithm for MCKP with running time
$\Oh(n+w_{\max}^4)$, based on upper-hull greedy anchors, a proximity bound, and
a subset-sum separation argument. The same dynamic program also yields
optimal values for $w_{\max}+1$ consecutive capacities. This windowed
viewpoint leads to applications in bounded-domain $(\max,+)$ convolution,
including deterministic network calculus. For instances with small price
ranges, the same approach yields a symmetric bound of $\Oh(n+p_{\max}^4)$. An open
problem is to match the best known small-item bounds for 0-1 Knapsack and
improve the dependence on $w_{\max}$ or $p_{\max}$ for MCKP.
For 0-1 Knapsack, stronger bounds in the small-item regime rely on refined
additive-combinatorial tools that exploit unrestricted mixing of items.
These techniques do not transfer cleanly to MCKP: the class constraint
prevents arbitrary combinations and breaks the additive structure that those
methods use. Closing this gap remains open.
In particular, the $\Oh(n+w_{\max}^3)$ bound for 0-1 Knapsack from
\cite{polak-etal21} relies on structural properties of unrestricted item sets
and does not seem to extend to the class-constrained setting.

\section*{Acknowledgments}
Supported by the National Science Centre (NCN) grant no. 2022/47/D/ST6/02184.

\bibliographystyle{plainurl}
\bibliography{mckp}

\end{document}